\documentclass[twocolumn,english,floatfix,aps,final]{revtex4}
\usepackage[T1]{fontenc}
\usepackage[latin9]{inputenc}
\usepackage{graphicx}

\usepackage{babel}

\begin{document}

\title{Impact of the relatively light fourth family neutrino on the Higgs
boson search}

\author{S.A. Cetin}

\email{scetin@dogus.edu.tr}

\affiliation{Do\u{g}u\c{s} University, Department of Physics, Istanbul, Turkey.}

\author{T. Çuhadar-Dönszelmann}

\email{tcuhadar@cern.ch}

\affiliation{University of Sheffield, Department of Physics and Astronomy, Sheffield,
U. K.}

\author{M. Sahin}

\email{m.sahin@etu.edu.tr}

\affiliation{TOBB University of Economics and Technology, Physics Division, Ankara,
Turkey.}

\author{S. Sultansoy}

\email{ssultansoy@etu.edu.tr}

\affiliation{TOBB University of Economics and Technology, Physics Division, Ankara,
Turkey.}

\affiliation{Institute of Physics, National Academy of Sciences, Baku, Azerbaijan}

\author{G. Unel}

\email{gokhan.unel@cern.ch}

\affiliation{Department of Physics and Astronomy, University of California at
Irvine, Irvine, CA, U.S.A.}
\begin{abstract}
The existence of a fourth fermion generation has mostly been considered
as a source of enhanced Higgs signals with respect to the 3 family
Standard Model predictions. However, a fourth Standard Model family
neutrino could cause the opposite situation. It is shown that relatively
light fourth family neutrino ($2m_{\nu_{4}}<m_{H}$) could drastically
change the interpretation of the search results for the Higgs boson,
especially if $m_{H}<170$ GeV. 
\end{abstract}
\maketitle
Discovery of the Higgs boson will complete confirmation of the Standard
Model (SM) basics. It is well known that the fourth SM family fermions
(for the remainder of this text, SM with three and four families will
be denoted as SM3 and SM4, respectively) have strong influence on
the Higgs boson properties \cite{Arik1,Jenni,Ginzburg,Sultansoy,Arik2,Arik3,Cakir,Arik4,Kribs,Arik5,Becerici}.
Especially, cross-section of the Higgs boson production via gluon
fusion at hadron colliders is essentially enhanced from about nine
times at the low Higgs boson mass values, to about four times at the
high values of the Higgs boson mass. If mass relations forbid decays
of the Higgs boson into fourth family fermions, this enhancement moves
to $gg\rightarrow H\rightarrow ZZ/WW$ channels, whereas the effect
becomes minor for $gg\rightarrow H\rightarrow\gamma\gamma$ channel
because of destructive contribution of the fourth family charged lepton
into $H\rightarrow\gamma\gamma$. However, situation could be changed
if Higgs boson decays into fourth family fermions are allowed, e.g.
for relatively light Higgs boson $\sigma(gg\rightarrow H)\times BR(H\rightarrow ZZ/WW)$
in SM4 may even be lower than in SM3 for certain values of the $\nu_{4}$
masses (see Tables I and II in \cite{Arik2}).

A review of the recent experimental results on the Higgs boson searches
is presented in \cite{Juste,Korytov}. In several decay channels,
the difference between the expectation and the observation reaches
$+2\sigma$, as reported by both LHC experiments. Two most important
differences are discussed below, as the remaining ones can be eliminated
as statistical fluctuation by comparing to the results from the other
channels. The discrepancies are in $H\rightarrow WW\rightarrow\ell\nu\ell\nu$
channel in the region $120<m_{H}<180$ GeV observed by both ATLAS
and CMS and in $H\rightarrow ZZ\rightarrow2\ell2q$ channel in the
vicinity of $m_{H}\approx500$ GeV as reported by CMS. The first excess
surpasses $2\sigma$ between $135$ and $150$ GeV as reported by
ATLAS and CMS result combination and CDF/D0 combination shows a similar
deviation in $H\rightarrow WW\rightarrow\ell\nu\ell\nu$, as well.
If the second excess is caused by the $H$ boson, then the required
cross-section becomes four times larger than the SM3 prediction \cite{Korytov}.
This is exactly the factor predicted by SM4 (see Table II in \cite{Arik2}).
In addition, ATLAS observed $2$ events around $500$ GeV in the {}``golden
mode''. Justification of this observation would again require a factor
of four to the production cross section. In the SM4 case, other channels
studied by ATLAS and CMS exclude $m_{H}$ around $500$ GeV at less
than $2\sigma$.

In this study, it is shown that the opening of the $H\rightarrow\nu_{4}\bar{\nu}_{4}$
channel could essentially modify situation in $135-150$ GeV region
mentioned above, whereas its effect is negligible at $500$ GeV since,
$WW$ and $ZZ$ modes are dominant for $m_{H}>170$ GeV independent
of $\nu_{4}$ mass. Nevertheless, if $\nu_{4}$ is unstable (i.e.
it decays within the detector) it could provide additional, so-called
{}``silver\textquotedblright \cite{Sultansoy2,Tulay}, channel for
Higgs discovery.

It is important to realise how the masses of the fourth generation
fermions affect Higgs properties. The dependence of the cross section
of the Higgs boson production via gluon fusion, on the masses of the
fourth generation quarks is illustrated in Figures 1 to 5 of \cite{Becerici}.
It can be seen that, the infinite mass limit of the fourth generation
quarks gives the most conservative enhancement to the production cross
section. This conservative enhancement has widely been used in most
of the related studies to stay on the pessimistic safe side. However,
making assumptions about the masses of the SM4 fermions might not
be the best approach when the decay properties of the Higgs boson
is under study.

The changes in the bosonic decay widths of the Higgs boson, e.g. $H\rightarrow gg,\,\gamma\gamma,\, Z\gamma$,
due to SM4 are almost independent of the SM4 fermion masses. However,
this is not the case for the decays of the Higgs boson into SM4 fermions,
especially the SM4 neutrino.

The decay width of $H\rightarrow\nu_{4}\bar{\nu}_{4}$ becomes very
dominant over the other decay widths of Higgs boson, especially at
relatively light SM4 neutrino mass. This property was demonstrated
earlier in \cite{Arik2} and also in recent papers \cite{Rozanov,Keung}.
The impact of such an SM4 neutrino can be observed in the change of
the production cross section and the branching ratio of the Higgs
boson in the following decay modes: $H\rightarrow WW,\, ZZ,\, ff$
and $H\rightarrow\gamma\gamma$. Here $f$ denotes a charged fermion.
The ratio of the $\sigma(gg\rightarrow H)\times BR(H\rightarrow X)$
in SM4 to SM3 case is the best parameter to demonstrate this effect.
In the remainder of this study, the masses of the charged SM4 fermions
are assumed to be heavy and $m_{\nu_{4}}$ is scanned from $45$ GeV
to $100$ GeV both for Dirac and Majorana cases as shown in Figures
1 to 8. Numerical calculations are performed using COMPHEP \cite{Comphep1,Comphep2},
HIGLU \cite{HIGLU} and HDECAY \cite{HDECAY} with appropriate modifications.

\begin{figure}
\includegraphics[scale=0.47]{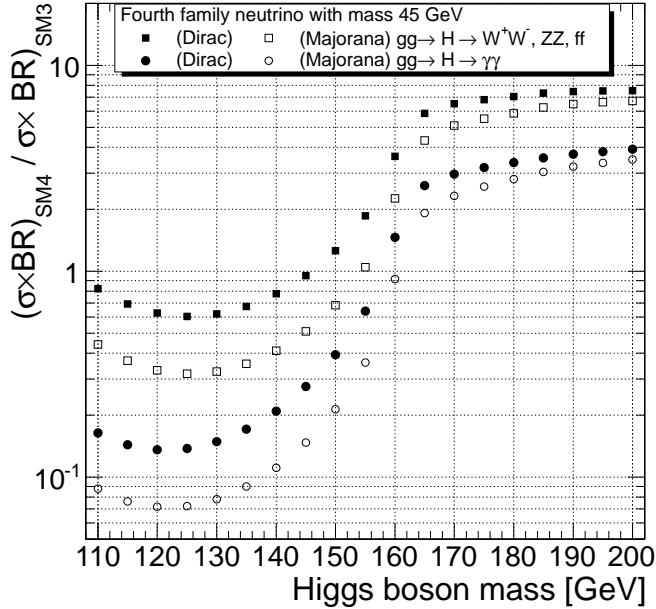}

\caption{Ratio of $\sigma(gg\rightarrow H)\times BR(H\rightarrow X)$ predicted
in SM4 to that in SM3 depending on the Higgs mass for $m_{\nu_{4}}=45$
GeV.}

\end{figure}

\begin{figure}
\includegraphics[scale=0.47]{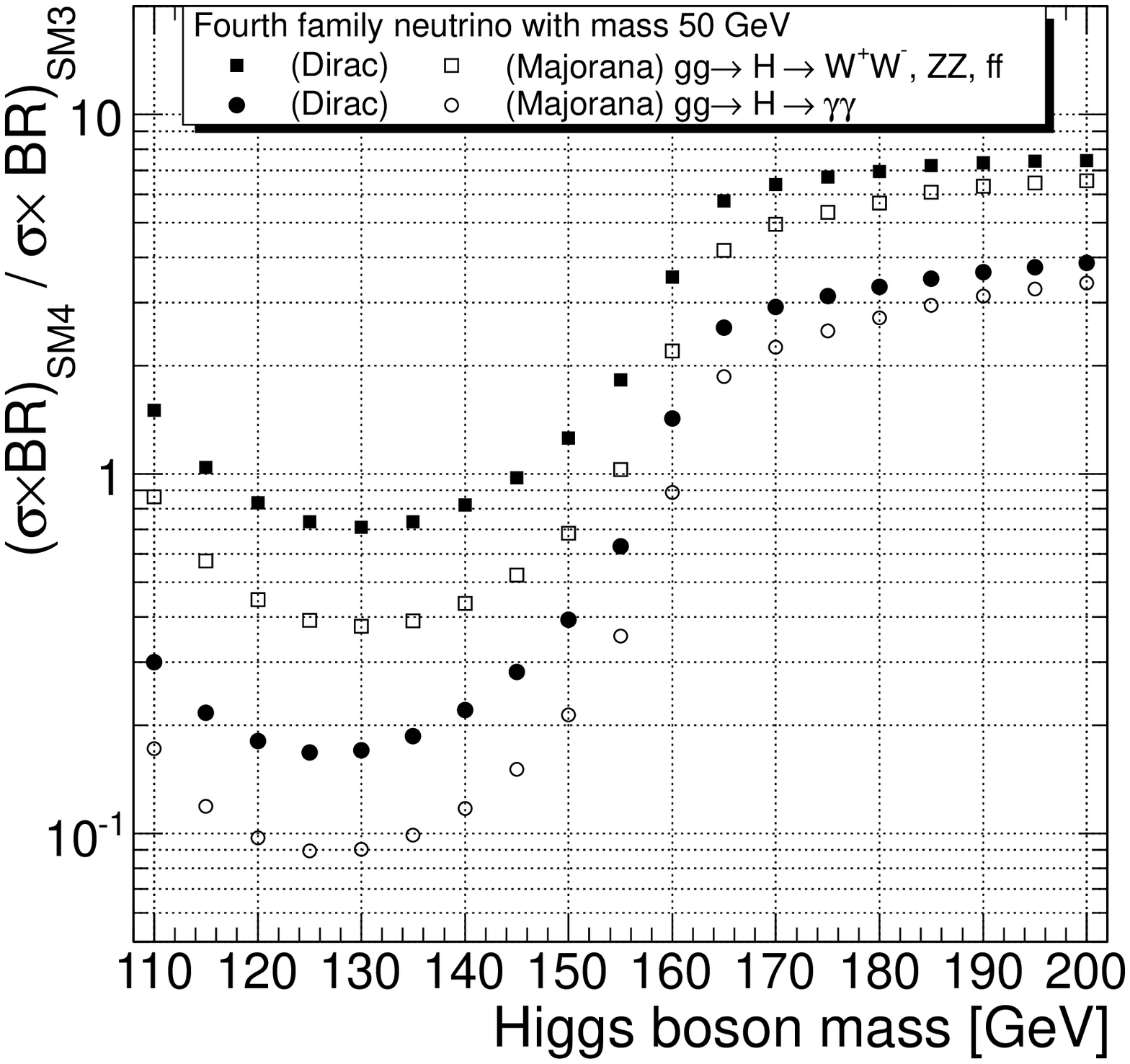}

\caption{Same as Fig. 1 but for $m_{\nu_{4}}=50$ GeV.}

\end{figure}

\begin{figure}
\includegraphics[scale=0.47]{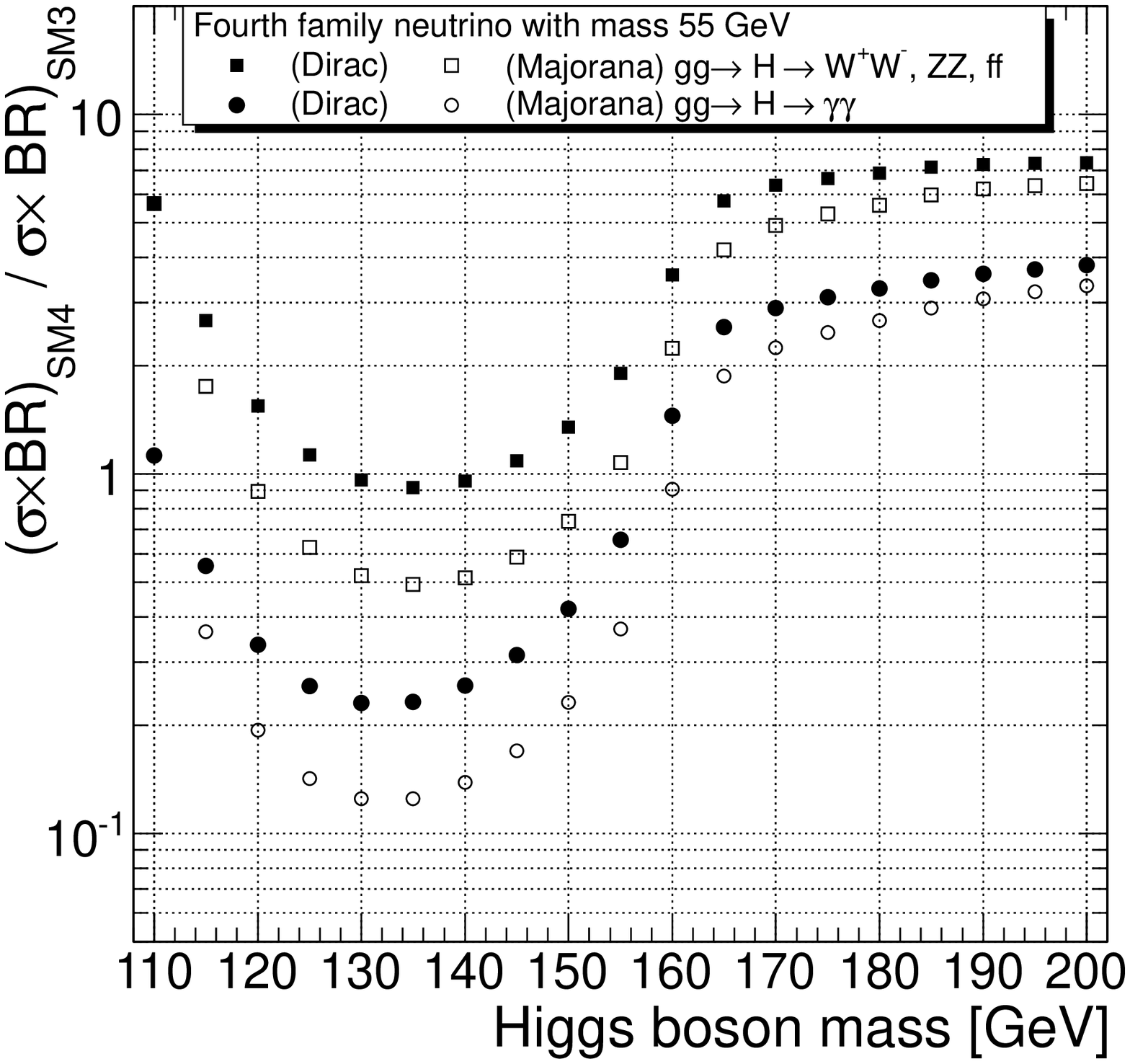}

\caption{Same as Fig. 1 but for $m_{\nu_{4}}=55$ GeV.}

\end{figure}

\begin{figure}
\includegraphics[scale=0.47]{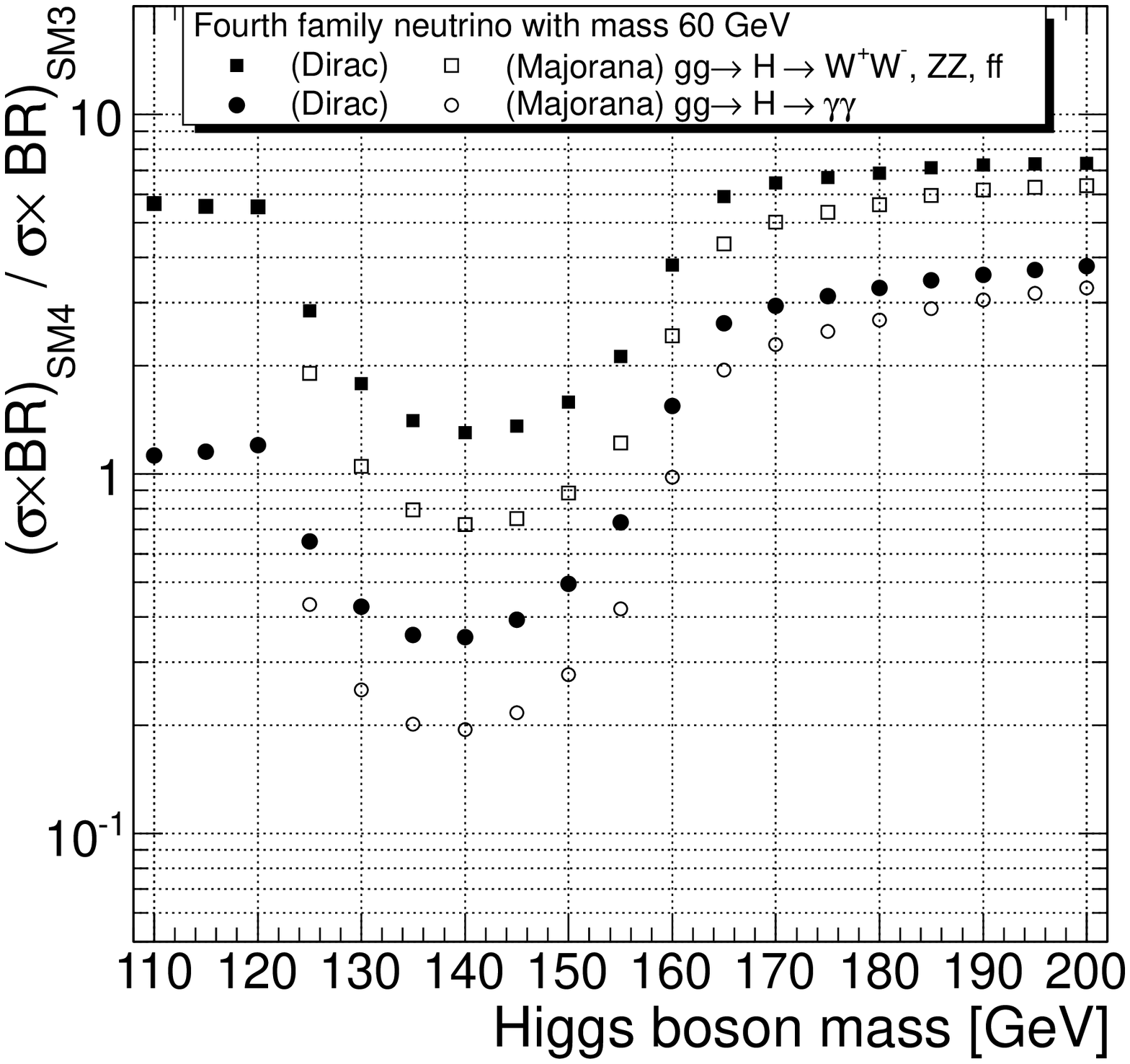}

\caption{Same as Fig. 1 but for $m_{\nu_{4}}=60$ GeV.}

\end{figure}

\begin{figure}
\includegraphics[scale=0.47]{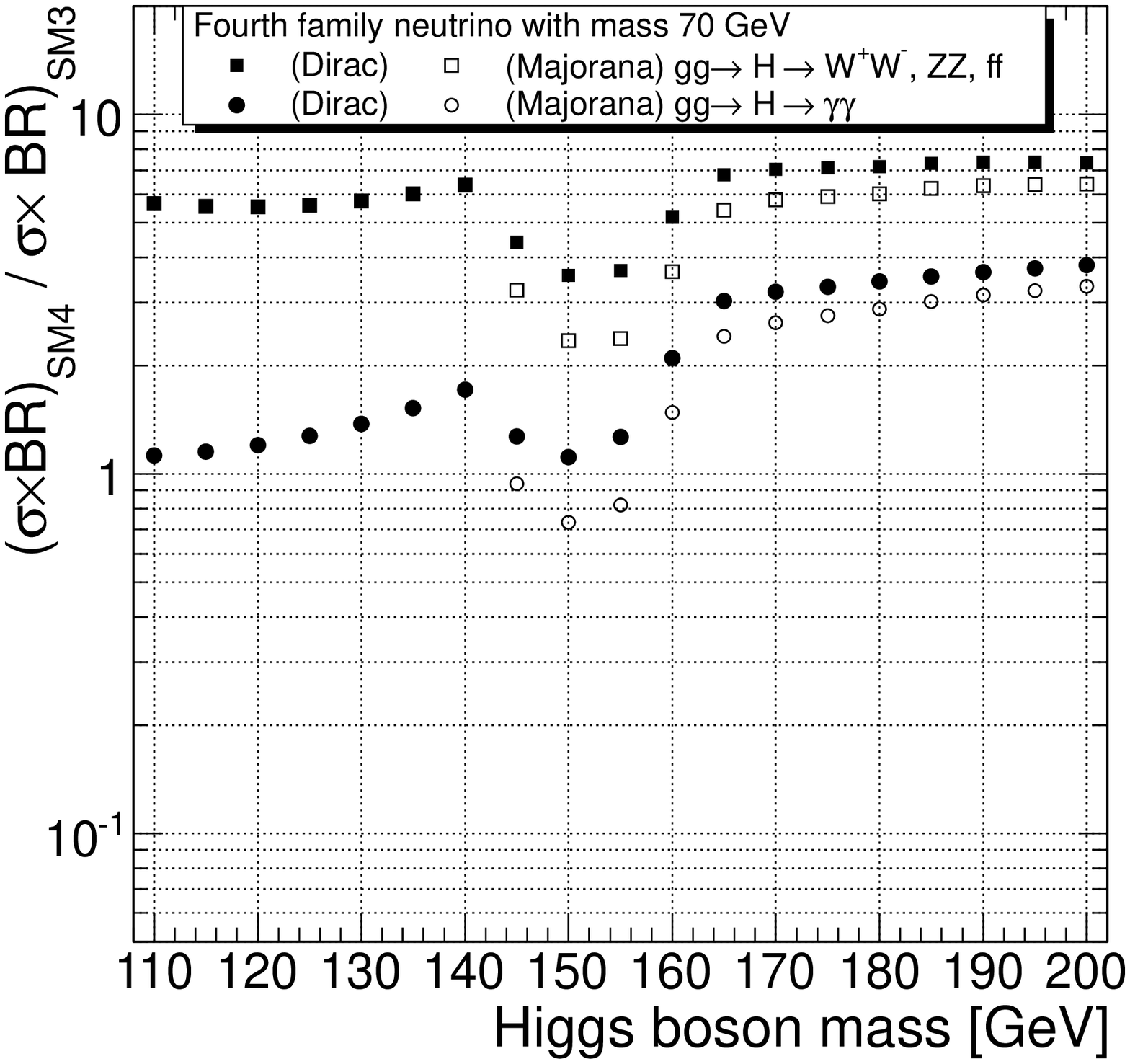}

\caption{Same as Fig. 1 but for $m_{\nu_{4}}=70$ GeV.}

\end{figure}

\begin{figure}
\includegraphics[scale=0.47]{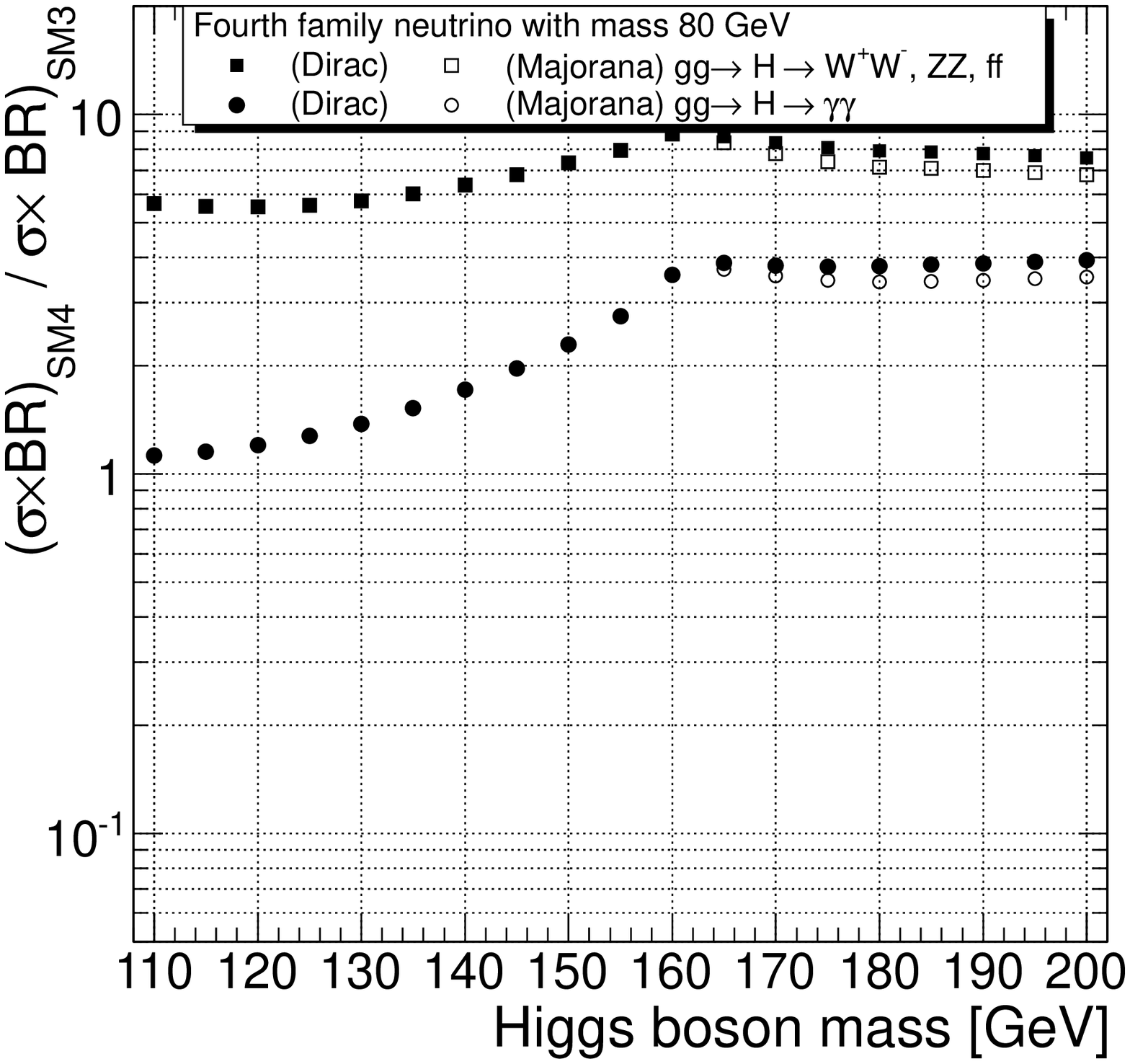}

\caption{Same as Fig. 1 but for $m_{\nu_{4}}=80$ GeV.}

\end{figure}

\begin{figure}
\includegraphics[scale=0.47]{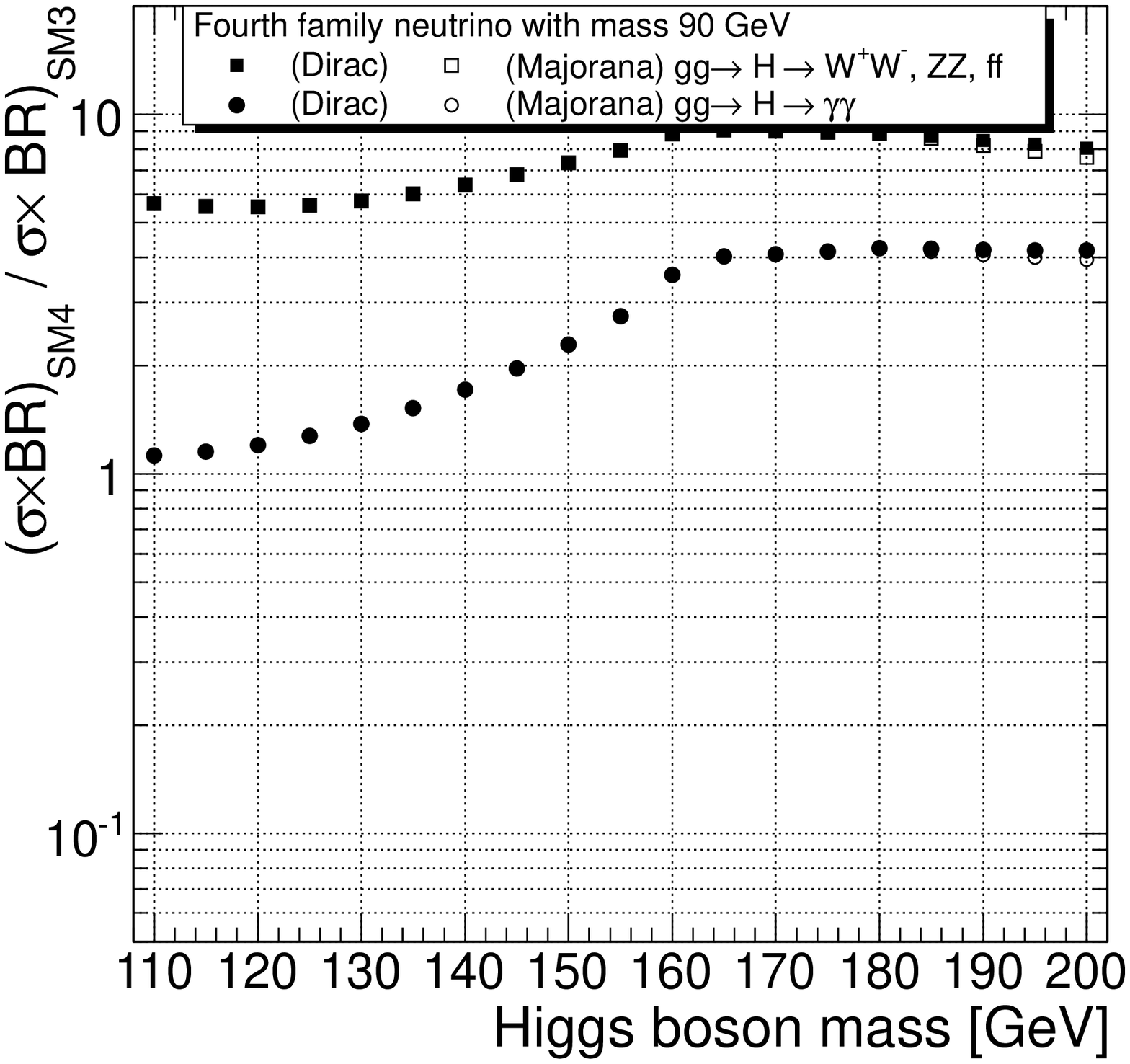}

\caption{Same as Fig. 1 but for $m_{\nu_{4}}=90$ GeV.}

\end{figure}

\begin{figure}[h]
 \includegraphics[scale=0.45]{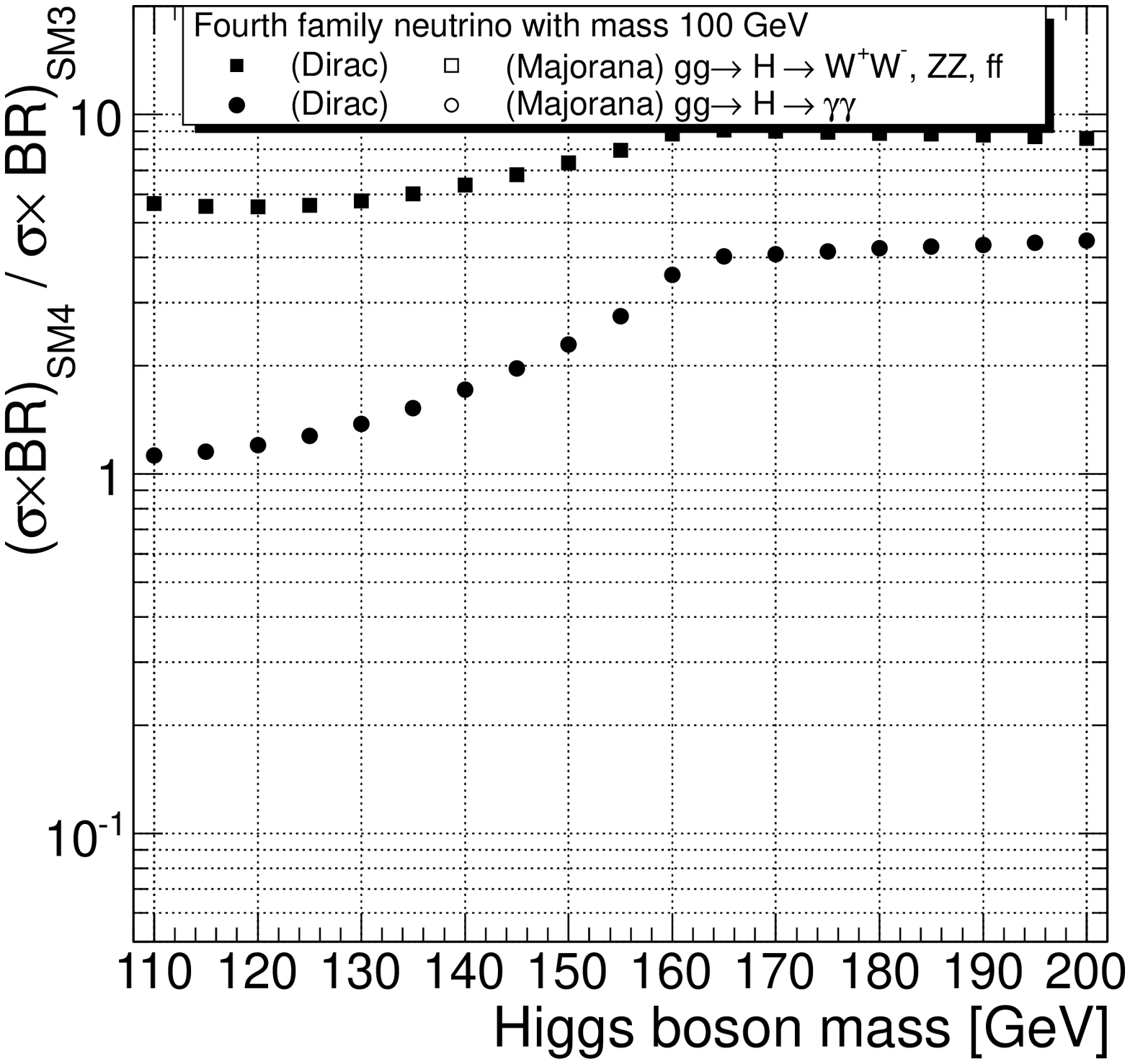}

\caption{Same as Fig. 1 but for $m_{\nu_{4}}=100$ GeV.}

\end{figure}

Experimental lower bounds on the SM4 neutrino mass are \cite{PDG}:
$45.0$ GeV ($90.3$ GeV) for stable (unstable) Dirac and $39.5$
GeV ($80.5$ GeV) for stable (unstable) Majorana neutrino Majorana
cases. In this context, stable means escaping the LEP detectors. For
the unstable Majorana neutrino mass, a lower limit of $62.1$ GeV
has been recently suggested \cite{Carpenter}.

\begin{figure}[ht!]
 \includegraphics[scale=0.45]{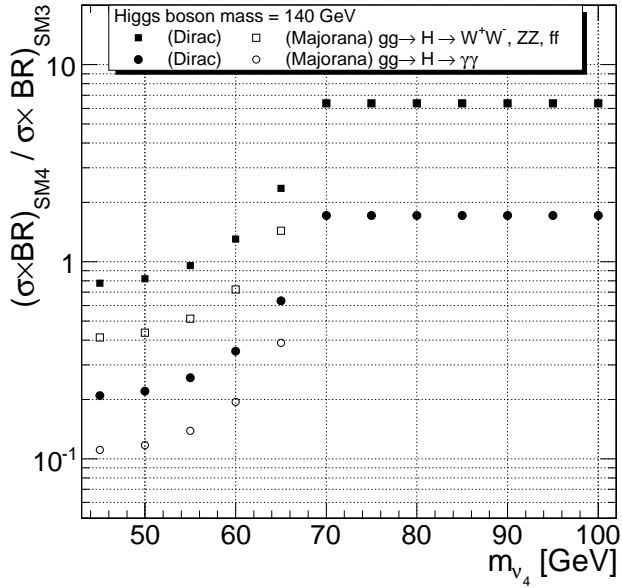}

\caption{Ratio of $\sigma(gg\rightarrow H)\times BR(H\rightarrow X)$ predicted
in SM4 to that in SM3 depending on $\nu_{4}$ mass for $m_{H}=140$
GeV.}

\end{figure}

The change in the branching ratio of $H\rightarrow WW,\, ZZ,\, ff$
channels is purely due to the change in the total decay width of Higgs
boson. In the case of $H\rightarrow\gamma\gamma$ the destructive
contribution of the charged SM4 lepton to the $H\rightarrow\gamma\gamma$
decay width results in a further decrease in its branching ratio.

Figures 1 through 8 show that SM4 case with neutrinos of masses up
to 70 GeV hides the Higgs boson with respect to SM3 expectations.
Hence, any limit put on the Higgs boson mass especially for $m_{H}<160$
GeV would need to be reconsidered in SM4 case. In other words {}``non
observation'' of a Higgs in this region would be interpreted as the
non exsistence of the Higgs in SM3 case, whereas this would not be
true in SM4 case.

It is seen that this ratio could essentially be below $1$ (in the
case of $H\rightarrow WW,\, ZZ,\, ff$) if $m_{\nu_{4}}$ is less
than $55\,(65)$ GeV in Dirac (Majorana) case. As a practical example
to visualize the influence of $\nu_{4}$ mass on the observable final
states, a $140$ GeV Higgs boson and heavy SM4 fermions with lighter
neutrino are considered in Figure 9. It is possible to use this Figure
in conjunction with the current experimental results (i.e. Fig. 4
from \cite{Juste,Korytov}) to make predictions on the type and mass
of SM4 neutrino. If the observed deviation around $140$ GeV is attributed
to an SM-like Higgs, then according to the bottom part of Fig. 4 in
\cite{Juste,Korytov} best fit signal strength is $\sigma/\sigma_{SM3}$
$\approx$ $0.5$ which corresponds to a Majorana neutrino with mass
$m{}_{\nu_{4}}=55$ GeV in Figure 9. Therefore, if studies with higher
luminosity reveal that the deviation is not a statistical fluctuation
but originates from a real Higgs boson, an SM4 with a \textquotedbl{}stable\textquotedbl{}
neutrino would yield a possible explanation for this unexpected result,
in addition to revealing the neutrino mass in an indirect way.

In conclusion, possible existence of the relatively light fourth family
neutrino requires re-interpretation of the ATLAS and CMS results in
the SM4 case. Namely, exclusion region for Higgs boson mass is reduced
from $120-600$ GeV to $160-500$ GeV. 
\begin{acknowledgments}
S. A. Cetin acknowledges the support from TAEK under the contract
2011TAEKCERN-A5.H2.P1.01-19 and also Turkish Academy of Sciences GEBIP
Programme. T. Çuhadar-Dönszelmann acknowledge support from the UK
Science and Technology Facilities Council. M. Sahin's work is supported
by TUBITAK BIDEB-2218 grant. S. Sultansoy acknowledges the support
from the Turkish State Planning Committee under the contract DPT2006K-120470
and from the Turkish Atomic Energy Authority (TAEK). G. Unel's work
is supported in part by U.S. Department of Energy Grant DE FG0291ER40679. 
\end{acknowledgments}


\vspace{33cm}


\vspace{-24cm}

\begin{thebibliography}{23}
\bibitem{Arik1} E. Arik \emph{et al}., ATLAS Internal Note ATL-PHYS-98-125,
1998.

\bibitem{Jenni} P. Jenni \emph{et al}., (ATLAS Collaboration), Report
No. CERN-LHCC-99-14/15, 1999, Sect. 18.2.

\bibitem{Ginzburg} I. F. Ginzburg, I. P. Ivanov, and A. Schiller,
Phys. Rev. D \textbf{60}, 095001 (1999).

\bibitem{Sultansoy} S. Sultansoy, hep-ex/0010037.

\bibitem{Arik2} E. Arik \emph{et al}., Phys. Rev. D \textbf{66},
033003 (2002).

\bibitem{Arik3} E. Arik \emph{et al}., Eur. Phys. J. C \textbf{26},
9 (2002).

\bibitem{Cakir} O. Cakir and S. Sultansoy, Phys. Rev. D \textbf{65},
013009 (2001).

\bibitem{Arik4} E. Arik \emph{et al}., Acta Phys. Pol. B \textbf{37},
2839 (2006).

\bibitem{Kribs} G. Kribs \emph{et al}., Phys. Rev. D \textbf{76},
075016 (2007).

\bibitem{Arik5} E. Arik, S. A. Cetin, and S. Sultansoy, Balkan Phys.
Lett. \textbf{15}, 1 (2007).

\bibitem{Becerici} N. Becerici Schmidt \emph{et al}., Eur. Phys.
J. C \textbf{66}, 119 (2010).

\bibitem{Juste} See ATLAS-CONF-2011-157 and the references therein.

\bibitem{Korytov} See CMS-PAS-HIG-11-023 and the references therein.

\bibitem{Sultansoy2} S. Sultansoy and G. Unel, Turk. J. Phys. \textbf{31,}
295 (2007); arXiv:0707.3266 {[}hep-ph].

\bibitem{Tulay} T. Çuhadar-Dönszelmann \emph{et al}., JHEP \textbf{10},
074 (2008).

\bibitem{Rozanov} A. N. Rozanov, M. I. Vysotsky, Phys. Lett. \textbf{B}
700, 313 (2011).

\bibitem{Keung} Wai-Yee Keung and Pedro Schwaller, JHEP 1106, 054
(2011).

\bibitem{PDG} PARTICLE DATA GROUP (Nakamura, K. \emph{et al}.) J.
Phys. G \textbf{37}, 075021 (2010).

\bibitem{Carpenter} L.M. Carpenter and A. Rajamaran, Phys. Rev. D
\textbf{82}, 114019 (2010).

\bibitem{Comphep1} A.Pukhov \emph{et al}., hep-ph/9908288.

\bibitem{Comphep2} E.Boos \emph{et al}., (CompHEP Collaboration),
Nucl. Instrum. Meth. A \textbf{534}, 250 (2004); arXiv:hep-ph/0403113.

\bibitem{HIGLU} M. Spira, Report DESY T-95-05 (1995), hep-ph/9510347.

\bibitem{HDECAY} A. Djouadi, J. Kalinowski, M. Spira, Comp. Phys.
Commun. \textbf{108}, 56 (1998); hep-ph/9704448. 
\end{thebibliography}
\end{document}